\documentclass[pra,showpacs,showkeys,amsfonts,amsmath,twocolumn]{revtex4}
\usepackage{bm}
\usepackage{graphicx}
\RequirePackage{mathptm}

\numberwithin{equation}{section}
\newcommand{\si}[1]{\sigma_{#1}}
\newcommand{\sa}[2]{\sigma_{#1}^{#2}}

\newcommand{\ro}{\varrho}
\newcommand{\ras}{\varrho_{\mr{as}}}
\newcommand{\la}{\lambda}

\newcommand{\al}{\alpha}
\newcommand{\be}{\beta}

\newcommand{\ga}[1]{\gamma_{#1}}
\newcommand{\Ga}[2]{\Gamma_{#1}^{#2}}
\newcommand{\Om}[1]{\Omega_{#1}}
\newcommand{\om}{\omega}
\newcommand{\te}{\vartheta}

\newcommand{\I}{\openone}
\newcommand{\conj}[1]{\overline{#1}}
\newcommand{\ket}[1]{|{#1}\rangle}
\newcommand{\bra}[1]{\langle {#1} |}

\newcommand{\C}{\mathbb C}

\newcommand{\tr}{\mathrm{tr}\,}

\newcommand{\mr}[1]{\mathrm{#1}}

\newcommand{\sr}[1]{\langle  {#1}\rangle}
\newcommand{\DS}{\displaystyle}
\newcommand{\rom}[1]{\rho_{#1}}
\newcommand{\bom}{\beta \omega}
\begin{document}
\title{Generation of Werner-like stationary states of two qubits in a thermal reservoir}
\author{Lech Jak{\'o}bczyk
\footnote{E-mail addres: ljak@ift.uni.wroc.pl}
}
 \affiliation{Institute of Theoretical Physics\\ University of
Wroc{\l}aw\\
Plac Maxa Borna 9, 50-204 Wroc{\l}aw, Poland}
\begin{abstract}
The dynamics of entanglement between two-level atoms immersed in the
common photon reservoir at finite temperature is investigated. It is shown
that in the regime of strong correlations there are nontrivial asymptotic
states which can be interpreted in terms of thermal generalization of Werner
states.
\end{abstract}
 \pacs{03.67.Mn, 03.65.Yz, 42.50.-p} \keywords{two-level atoms, entanglement production, thermal reservoir}
\maketitle
\section{Introduction}
Entanglement generation by the indirect interaction between
otherwise decoupled systems has been discussed in the literature
mainly in the case of two-level atoms interacting with the common
vacuum.  The idea that dissipation can create rather then destroy
entanglement in some systems, was put forward in several publications
\cite{Plenio, Kim, SM, PH}. In particular, the effect of spontaneous
emission on destruction and production of entanglement was discussed
\cite{J, FT1, FT2, JJ1}. When the two atoms are separated by a distance small
compared to the radiation wave length, the collective properties of
two-atomic systems can alter the decay process compared with the single atom.
It was already shown by Dicke \cite{Di} that there are states with enhanced
emission rates (superradiant states) and such that the emission rate is reduced
(subradiant states). (In the case of multi-atomic systems, similar collective
effects described by Tavis-Cummings interaction \cite{TC} can generate collective
multiqubit entanglement \cite{RSR}). When the emission rate is reduced,
two-atom system can decohere slower
compared with individual atoms and some amount of initial entanglement can be
preserved or even created by the indirect interaction between atoms.
The analogous effect of production of entanglement
was studied in a system of two-level atoms interacting with a squeezed vacuum.
In that case, the squeezed vacuum is a source of nonclassical correlations
which are essential for creation of entanglement \cite{TF}. The general condition
under which entanglement can be induced by interaction with environment was also 
discovered \cite{BFP}.
\par
The case of two atoms immersed in a common thermal reservoir was
also investigated \cite{BF, ZY, L}. In particular, Benatti and
Floreanini \cite{BF} have discussed  the dynamics of two independent
atoms interacting with the reservoir of scalar particles at finite
temperature and found the interesting behavior of the system.  If
the atoms are at finite separation, there is  a temperature of the
reservoir below which entanglement generation occurs. Moreover, for
the vanishing separation, the entanglement thus generated persists
in the asymptotic state.
\par
In the present paper, we study the similar model but from a
different perspective. We consider the system of two-level atoms
interacting with the photon reservoir at fixed temperature. As in
the vacuum case, the collective properties of the atomic system can
alter the decay process compared with the single atom. There are
states with enhanced emission rates   and such that the emission
rate is reduced. The important example of the latter  is the singlet
state $\ket{a}$ i.e. antisymmetric superposition constructed from
energy levels of considered atoms. As we show, in the regime of
strong correlations the singlet state is decoupled from the
environment and therefore is stable. In our considerations we focus
on this case and investigate the stationary asymptotic states of the
system. In turns out that the asymptotic states are parametrized by
the fidelity $F$ of the initial state $\ro$ with respect to the
state $\ket{a}$ (or the overlap of $\ro$ with singlet $\ket{a}$) and
the temperature $T$ of the photon reservoir. We  identify the
asymptotic states as the \textit{thermal generalization of Werner
states} i.e. mixture of singlet state and Gibbs equilibrium state at
the temperature $T$ with the appropriate probability. (The standard
Werner states \cite{W} are recovered in the limit of infinite
temperature). Depending on the initial fidelity some of the
asymptotic states are entangled.  We calculate the amount of the
asymptotic entanglement  using  a concurrence as its measure. We
also show that if the initial fidelity is greater or equal to $1/2$,
then for every finite temperature of the reservoir the asymptotic
entanglement is non-zero. On the other hand, for fidelity less then
$1/2$ there is \textit{the critical temperature} below  which the
asymptotic states are entangled whereas for higher temperatures all
stationary states are separable.
\par
Since the initial states with the same fidelity can be separable or
entangled, the dynamics given by the interaction with the reservoir
manifests differently with respect to the entanglement properties of
the system. It can create entanglement when the initial  states are
separable, disentangle initially entangled states, preserve some
part of initial entanglement or even increase the initial
entanglement of some states. We show on specific examples  that this
behavior of the dynamics really occurs. In particular, pure product
states with orthogonal factor vectors become entangled in every
temperature, whereas for other product states there is the critical
temperature for creation of entanglement. Similar phenomenon can be
observed if the initial state is mixed. For example in the case of
Gibbs equilibrium state at the temperature $T_{0}$ which differs
from the temperature $T$ of the reservoir, the creation of
asymptotic entanglement happens if $T_{0}$ is much higher then $T$.
On the other hand, entangled initial states with maximal amount of
entanglement can preserve some initial entanglement or can
disentangle completely. It is worth to stress that this different
behavior with respect to the thermal noise can happen for locally
equivalent initial states, so by performing only local operations
one can protect as much entanglement as possible. When the initial
states are not maximally entangled, then   the thermal noise can in
some cases create  entanglement which adds to the initial one,
although during the evolution the purity is decreasing.

\section{Two qubits dynamics}
We start with the sketch of the derivation of the dynamical equation
describing the evolution of the system of atoms immersed in the
thermal reservoir (for details see e.g. \cite{FT}).
\par
Consider two-level atoms $A$ and $B$ with ground states
$\ket{0}_{j}$ and excited states $\ket{1}_{j}$ ($j=A,\, B$),
interacting with photon reservoir at temperature $T$. The dynamics
of the combined system consisting of the atoms and the quantum
electromagnetic field is given by the Hamiltonian
$$
H_{\mr{SF}}=H_{0}+H_{I}
$$
where $H_{0}$ is the sum of free Hamiltonians of the atoms and the
field and $H_{I}$ describes the interaction between the atoms and
photons in the electric dipole approximation. Since we are
interested in the dynamics of the system of atoms, we take the
partial trace over field variables and find that the reduced density
matrix of the atomic system satisfies some integro-differential
equation which can be simplified by employing the Born approximation
(the interaction between the atoms and the field is so weak that
there is no back reaction effect of the atoms on the field). Under
this approximation, the evolution of the density matrix depends on
the first- and second-order correlation functions of the field
operators in the thermal equilibrium state at temperature $T$.
Employing further the rotating-wave approximation in which we ignore
all terms oscillating at higher frequencies and assuming that the
correlation time of the photon reservoir is short (the Markov
approximation), we arrive at the result that the influence of
reservoir on the system of atoms can be described by dynamical
semi-group \cite {AL} with Lindblad generator
$L=-i\,[H,\,\cdot]+L_{\mr{D}}$, where
\begin{equation}
H=\frac{\om}{2}\sum\limits_{j=A,B}\sa{3}{j}+\sum\limits_{j,\,k=A,B\atop
j\neq k} \Om{jk}\sa{+}{j}\sa{-}{k}\label{hamilt}
\end{equation}
and
\begin{equation}
\begin{split}
L_{\mr{D}}\ro=&\frac{1}{2}\sum\limits_{j,\,k=A,\,B}\Ga{jk}{\downarrow}\,\left(2\sa{-}{j}\ro\sa{+}{k}-\sa{+}{k}\sa{-}{j}\ro-\ro\sa{+}{k}\sa{-}{j}\right)\\
&\hspace*{9mm}
+\Ga{jk}{\uparrow}\,\left(2\sa{+}{j}\ro\sa{-}{k}-\sa{-}{k}\sa{+}{j}\ro-\ro\sa{-}{k}\sa{+}{j}\right)
\end{split}\label{dyss}
\end{equation}
Here
$$
\sa{\pm}{A}=\sigma_{\pm}\otimes\I,\quad
\sa{\pm}{B}=\I\otimes\sigma_{\pm},\quad
\sa{3}{A}=\sigma_{3}\otimes\I,\quad \sa{3}{B}=\I\otimes\sigma_{3}
$$
In the Hamiltonian (\ref{hamilt}), $\omega$ is the frequency of the
transition $\ket{0}_{j}\to\ket{1}_{j}$ ($j=A,\, B$) and
$\Om{AB}=\Om{BA}=\Omega$ describes interatomic coupling by the
dipole-dipole interaction. On the other hand, dissipative dynamics
is given by the generator (\ref{dyss}) with
$$
\Ga{jk}{\downarrow}=\ga{jk}\,(1+\sr{n}),\quad
\Ga{jk}{\uparrow}=\ga{jk}\,\sr{n}\label{Ga}
$$
where
$$
\sr{n}=\frac{e^{-\be\om}}{1-e^{-\be\om}},\quad \be=\frac{1}{T}
$$
is the mean number of photons, and
\begin{equation}
\ga{AA}=\ga{BB}=\ga{0},\quad \ga{AB}=\ga{BA}=\gamma\label{ga}
\end{equation}
In the above equalities, $\ga{0}$ is the single atom spontaneous
emission rate, and $\gamma=G\ga{0}$ is the collective damping
constant. In the model considered, $G$ is the function of the
interatomic distance $R$, and $G$ is small for large separation of
atoms. On the other hand, $G\to 1$ when $R$ is small.
\par
The master equation
\begin{equation}
\frac{d\ro}{dt}=L\ro\label{me}
\end{equation}
giving the time evolution of a density matrix of the system of
two-level atoms can be used to obtain the equations for its matrix
elements with respect to some basis. To simplify  the calculations
one can work in the basis of collective states in the Hilbert space
$\C^{2}\otimes\C^{2}$ \cite{FT}, given by product vectors
\begin{equation}
\ket{e}=\ket{1}_{A}\otimes\ket{1}_{B},\quad
\ket{g}=\ket{0}_{A}\otimes\ket{0}_{B}
\end{equation}
symmetric superposition
\begin{equation}
\ket{s}=\frac{1}{\sqrt{2}}\left(\ket{0}_{A}\otimes\ket{1}_{B}+\ket{1}_{A}\otimes\ket{0}_{B}\right)
\end{equation}
and antisymmetric superposition
\begin{equation}
\ket{a}=\frac{1}{\sqrt{2}}\left(\ket{1}_{A}\otimes\ket{0}_{B}-\ket{0}_{A}\otimes\ket{1}_{B}\right)
\label{a}
\end{equation}
In the basis of collective states, two-atom system can be treated as
a single four-level system with ground state $\ket{g}$, excited
state $\ket{e}$ and two intermediate states $\ket{s}$ and $\ket{a}$.
From (\ref{me}) it follows that the matrix elements of the state
$\ro$ with respect to the basis $\ket{e},\, \ket{s},\,\ket{a},\,
\ket{g}$ satisfy the  equations which can be grouped into decoupled
systems of differential equations. So for diagonal matrix elements
we obtain
\begin{equation}
\begin{split}
\frac{d\rom{ee}}{dt}&=(\ga{0}-\gamma)\sr{n}\rom{aa}-2\ga{0}(1+\sr{n})\rom{ee}+(\ga{0}+\gamma)\sr{n}\rom{ss}\\
\frac{d\rom{ss}}{dt}&=-(\ga{0}+\gamma)[(1+\sr{n})\rom{ee}-(1+2\sr{n})\rom{ss}+\sr{n}\rom{gg}]\\
\frac{d\rom{aa}}{dt}&=-(\ga{0}-\gamma)[(1+2\sr{n})\rom{aa}-(1+\sr{n})\rom{ee}-\sr{n}\rom{gg}]\\
\frac{d\rom{gg}}{dt}&=(\ga{0}-\gamma)(1+\sr{n})\rom{aa}+(\ga{0}+\gamma)\rom{ss}-2\ga{0}\sr{n}\rom{gg}
\end{split}\label{diag}
\end{equation}
On the other hand, the elements $\rom{es}$ and $\rom{sg}$ are
connected by the equations
\begin{equation}
\begin{split}
\frac{d\rom{es}}{dt}&=(\ga{0}+\gamma)\sr{n}\rom{sg}-\frac{1}{2}
[(1+2\sr{n})\gamma\\
&\hspace*{4mm}+(3+4\sr{n})\ga{0}-2i(\om-\Omega)]\rom{es}\\
\frac{d\rom{sg}}{dt}&=(\ga{0}+\gamma)(1+\sr{n})\rom{es}-\frac{1}{2}[(1+2\sr{n})\gamma\\
&\hspace*{4mm}+(1+4\sr{n})\ga{0}+ 2i(\om+\Omega)]\rom{sg}
\end{split}\label{es}
\end{equation}
and similarly, the elements $\rom{ea}$ and $\rom{ag}$ satisfy
\begin{equation}
\begin{split}
\frac{d\rom{ea}}{dt}&=-(\ga{0}-\gamma)\sr{n}\rom{ag}+\frac{1}{2}[(1+2\sr{n})\gamma\\
&\hspace*{4mm}-(3+4\sr{n})\ga{0} -2i(\om+\Omega)]\rom{ea}\\
\frac{d\rom{ag}}{dt}&=-(\ga{0}-\gamma)(1+\sr{n})\rom{ea}-\frac{1}{2}[(1+4\sr{n})\ga{0}\\
&\hspace*{4mm}-(1+2\sr{n})\gamma +2i(\om-\Omega)]\rom{ag}
\end{split}\label{ea}
\end{equation}
Finally
\begin{equation}
\frac{d\rom{eg}}{dt}=-[(1+2\sr{n})\ga{0}+2i\om]\rom{eg}\label{eg}
\end{equation}
and
\begin{equation}
\frac{d\rom{sa}}{dt}=-[(1+2\sr{n})\ga{0}+2i\Omega]\rom{sa}\label{sa}
\end{equation}
The equations for the remaining matrix elements can be obtained by
using hermiticity of $\ro$.
\par
 From the equations (\ref{diag}) it follows that similarly as in the zero temperature case (see e.g. \cite{FT}),
 the system of atoms prepared in the symmetric state $\ket{s}$ decays with enhanced
rate $\ga{0}+\gamma$, whereas antisymmetric initial state $\ket{a}$
leads to the reduced rate $\ga{0}-\gamma$. In the limiting case of
strongly correlated atoms we can put $\gamma=\ga{0}$, so the state
$\ket{a}$ is completely decoupled from the photon reservoir. One can
also check that the master equation (\ref{me}) describes two types
of time evolution of the atomic system, depending on the relation
between $\gamma$ and $\ga{0}$. When $\gamma<\ga{0}$, there is a
unique asymptotic state of the system, which is the Gibbs state
\begin{equation}
\ro_{\be}=e^{-\be H_{0}}\big/\tr\,e^{-\be H_{0}},\quad
H_{0}=\frac{\om}{2}\sum\limits_{j=A,B}\sa{3}{j}\label{Gibbs}
\end{equation}
The state (\ref{Gibbs}) is separable and describes thermal
equilibrium of atoms interacting with photon reservoir. In the
regime of strong correlations, $\gamma=\ga{0}$ and  we show that
there are nontrivial asymptotic stationary states which can be
parametrized by matrix elements $\rom{aa}$ of the initial state.
\section{Strongly correlated qubits and nontrivial asymptotic states}
When $\gamma=\ga{0}$, equations (\ref{diag}) - (\ref{sa}) simplify
and one can check that the solutions of (\ref{es}) - (\ref{sa})
asymptotically vanish, so the only contribution to the stationary
states $\ras$ comes from the matrix elements $\rom{aa},\,
\rom{ss},\rom{ee}$ and $\rom{gg}$. Notice that
$$
\frac{d\rom{aa}}{dt}=0,\quad\text{so}\quad \rom{aa}(t)=\rom{aa}(0)=F
$$
where
$$
 F=\bra{a}\ro\ket{a}
$$
is the \textit{fidelity} of the initial state $\ro$ with respect to
the singlet state $\ket{a}$. Hence $ (\ras)_{aa}=F$  and after a long elementary
calculation, we obtain that
\begin{equation}
\begin{split}
(\ras)_{ee}&=\frac{e^{-2 \bom}}{u(\be)}\;(1-F)\\
(\ras)_{ss}&=\frac{e^{-\bom}}{u(\be)}\;(1-F)\\
(\ras)_{gg}&=\frac{1}{u(\be)}\;(1-F)
\end{split}\label{asymp}
\end{equation}
where
$$
u(\be)=1+e^{-\bom}+e^{-2 \bom}
$$
In the canonical basis
$$
\ket{1}_{A}\otimes\ket{1}_{B},\;\ket{1}_{A}\otimes\ket{0}_{B},\;
\ket{0}_{A}\otimes\ket{1}_{B},\; \ket{0}_{A}\otimes\ket{0}_{B}
$$
the non-zero matrix elements of the asymptotic state reads
\begin{equation}
\begin{split}
(\ras)_{11}&=\frac{e^{-2\bom}}{u(\be)}\;(1-F)\\
(\ras)_{22}&=\frac{e^{-\bom}}{2u(\be)}\;(1-F)+\frac{F}{2}\\
(\ras)_{23}&=\frac{e^{-\bom}}{2u(\be)}\;(1-F)-\frac{F}{2}\\
(\ras)_{44}&=\frac{1}{u(\be)}\;(1-F)
\end{split}\label{kanasymp}
\end{equation}
and $(\ras)_{33}=(\ras)_{22}$.
\par
The asymptotic state $\ras$ defined by (\ref{kanasymp}) exists for
any initial state and for the fixed temperature of the photon
reservoir it depends only on the initial fidelity i.e.
$\ras=\ras(F)$. If we define \textit{the threshold fidelity}
$F_{\be}$ by
$$
F_{\be}=\frac{e^{-\bom}}{(1+e^{-\bom})^{2}}
$$
then one can check that:\\[2mm]
(1) $\ras(F_{\be})$ equals the the Gibbs state $\ro_{\be}$,\\[2mm]
(2) for $F>F_{\be}$, $\ras (F)$ equals to the \textit{thermal
generalization of the Werner state}
\begin{equation}
W_{\be}=(1-p)\,\ro_{\be}+p\,\ket{a}\bra{a}\label{thermalW}
\end{equation}
where the mixing probability $p$ depends  on the fidelity of the initial state and temperature of the reservoir\,:
\begin{equation}
p=\frac{(1+e^{-\bom})^{2}\,F-e^{-\bom}}{u(\be)}\label{p}
\end{equation}
(3) for $F<F_{\be}$, the state $\ras$ cannot be expressed as the
Werner state (\ref{thermalW})\\

 Notice that in the limit $\be\to 0$ (or $T\to \infty$) we obtain
the standard Werner state \cite{W}, hence the dynamical generation
of such states occurs due to the interaction with the environment
with maximal noise \cite{JJ}. The entanglement properties of the
asymptotic state will be discussed in the next section.
\section{Asymptotic entanglement}
We start with the characterization of entanglement of the thermal
Werner state $W_{\be}$. The simplest way to do this is to use
Wootters concurrence \cite{Woot} defined for any two-qubit state
$\ro$ as
\begin{equation}
C(\ro)=\max
\left(\,0,\sqrt{\la_{1}}-\sqrt{\la_{2}}-\sqrt{\la_{3}}-\sqrt{\la_{4}}\,\right)
\end{equation}
where $\la_{1}>\la_{2}>\la_{3}>\la_{4}$ are the eigenvalues of the
matrix $\ro\widetilde{\ro}$ with $\widetilde{\ro}$ given by
$$
\widetilde{\ro}=\si{2}\otimes\si{2}\,\conj{\ro}\,\si{2}\otimes\si{2}
$$
where $\conj{\ro}$ denotes complex conjugation of the matrix $\ro$.
By a direct calculation one obtains  that for the states
\begin{equation}
\ro=\begin{pmatrix}\ro_{11}&0&0&0\\
0&\ro_{22}&\ro_{23}&0\\
0&\ro_{32}&\ro_{33}&0\\
0&0&0&\ro_{44}\end{pmatrix}\label{states}
\end{equation}
the concurrence is given by the simple function
\begin{equation}
C(\ro)=\max\,\left(\,0,\;2\,(|\ro_{23}|-\sqrt{\ro_{11}\ro_{44}})\;\right)\label{conc}
\end{equation}
Since the states (\ref{thermalW}) are of the form (\ref{states}),
$C(W_{\be})$ reads
\begin{equation}
C(W_{\be})=\max\,\left(\,
0,\;p-\frac{2e^{-\bom}}{(1+e^{-\bom})^{2}}\,(1-p)\,\right)\label{Wcomc}
\end{equation}
so for
$$
p>p_{0}=\frac{2e^{-\bom}}{1+4e^{-\bom}+e^{-2\bom}}
$$
thermal Werner states are entangled, and for $p\leq p_{0}$, those
states are separable. Combining this result with the formula
(\ref{p}), we obtain the concurrence of the asymptotic states (see
also \cite{BF, L})
\begin{equation}
C_{\mr{as}}=\max\,\left(\,0,\,F-\frac{3}{1+2\cosh\bom}\,(1-F)\;\right)\label{aconc}
\end{equation}
Observe that for the fixed temperature of the reservoir the
asymptotic entanglement depends only on the fidelity of the initial
state. Moreover, this entanglement is non-zero for all initial
states with fidelity satisfying
\begin{equation}
F>F_{0}=\frac{3}{4+2\cosh\bom}\label{F}
\end{equation}
Obviously, all asymptotic states with fidelity less then threshold
value $F_{\be}$ are separable.
\par
We can also consider the interesting problem of temperature
dependence of the asymptotic entanglement. Notice that for every
$\be> 0$ and $F\geq 1/2$
$$
F-\frac{3}{1+2\cosh\bom}\,(1-F)>0
$$
So the interaction of the atomic system with the reservoir at any
finite temperature brings all initial states  with the fidelity $F$
greater or equal to $1/2$ into the stationary  entangled states.
 On the other hand, when the fidelity is smaller then $1/2$, there is
the \textit{critical temperature}  i.e such temperature $T_{\mr{c}}$
that if $T<T_{\mr{c}}$ the  asymptotic states are entangled whereas
for $T\geq T_{\mr{c}}$ the asymptotic states corresponding to the
same initial fidelity are separable. Simple calculation shows that
\begin{equation}
\frac{T_{\mr{c}}}{\om}=\varphi(F)\label{Tc}
\end{equation}
where for $F<1/2$
\begin{equation}
\varphi
(F)=\left[\ln\left(\frac{\sqrt{3}\sqrt{3-8F+4F^{2}}+3-4F}{2F}\right)\right]^{-1}\label{fTc}
\end{equation}
\par
Since the states with the same fidelity can be separable or
entangled, we  expect that the system will behave differently
depending on the initial conditions. More precisely, it can happen
that unentangled atoms become entangled during the evolution and
initially entangled states disentangle or remain entangled. Next we
show that the above possibilities actually occur.
\subsection{Separable initial states}
When the initial state is the pure product state
\begin{equation}
\ket{\Psi}=\ket{\varphi}\otimes\ket{\psi},\quad
\ket{\varphi},\ket{\psi}\in \C^{2}\label{puresep}
\end{equation}
then its fidelity is given by
\begin{equation}
F=\frac{1}{2}\left(1-|\bra{\varphi}\psi\,\rangle|^{2}\,\right)\label{prodfid}
\end{equation}
If we denote $|\bra{\varphi}\psi\,\rangle|=\al$,  the concurrence of
the asymptotic state corresponding to (\ref{puresep}) can be
computed from the formula
\begin{equation}
C_{\mr{as}}(\al)=\max\,\left(\,0,\,
\frac{(1-\al^{2})\cosh\bom-1-2\al^{2}}{1+2\cosh\bom}\,\right)\label{psias}
\end{equation}
\begin{figure}[h]
\centering
{\includegraphics[height=54mm]{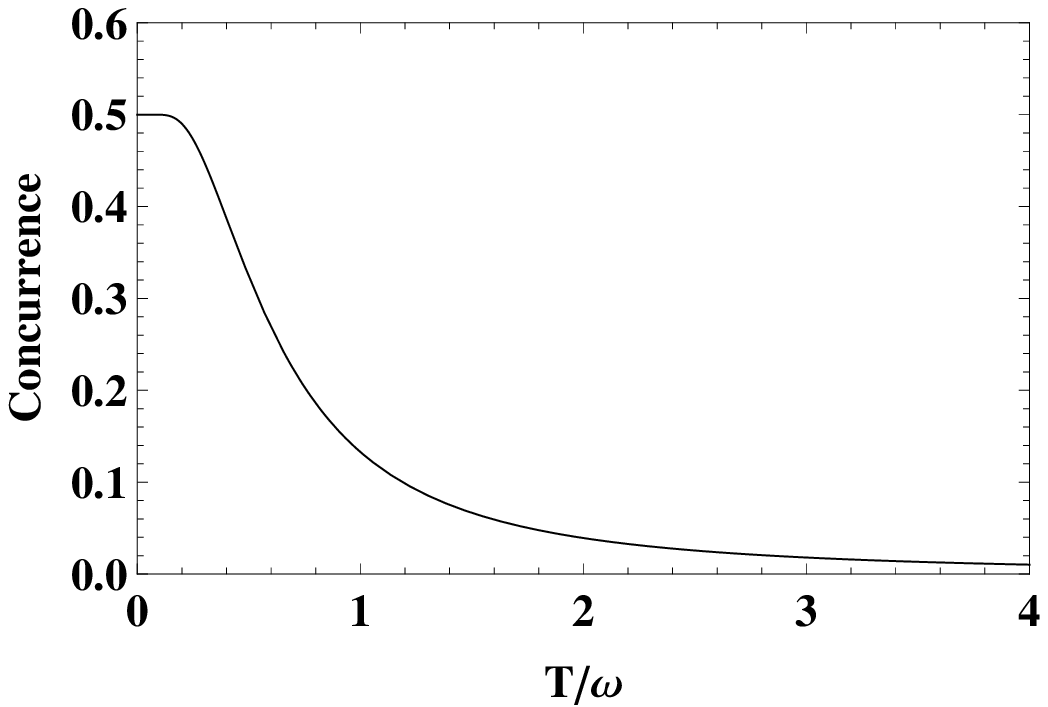}}\caption{$C_{\mr{as}}(\al)$
as a function of temperature for the  state (\ref{puresep}) with
$\al=0$.}
\end{figure}
In the case when the vectors $\varphi$ and $\psi$ are orthogonal
(for example $\varphi$ is the excited state of the atom $A$ and
$\psi$ is the ground state of the atom $B$), $F$ is maximal and for
every finite temperature the concurrence of the asymptotic state is
non-zero. So the interaction with the photon reservoir creates the
stationary entanglement between initially unentangled atoms. The
amount of this entanglement is maximal for the zero temperature and
decreases asymptotically to zero when the temperature of the
reservoir increases to infinity(FIG. 1). Notice also that for
$\al=1,\; C_{\mr{as}}(\al)$ trivially vanish. In the general case
when $0<\al<1$, the creation of the stationary entanglement is
possible only for the temperatures of the reservoir below the
critical temperature, which in that case is given by the formula
\begin{equation}
\frac{T_{\mr{c}}}{\om}=\left[\,\ln\left(\frac{\sqrt{3}\sqrt{2\al^{2}+\al^{4}}+1
+2\al^{2}}{1-\al^{2}}\right)\,\right]^{-1} \label{Tca}
\end{equation}
Since $T_{\mr{c}}$ goes to infinity when $\al$ goes to zero,  the
critical temperature can be high for some initial states, but the
created entanglement is always maximal for zero temperature and
vanishes for $T=T_{\mr{c}}$ (FIG. 2).
\begin{figure}[t]
\centering
{\includegraphics[height=54mm]{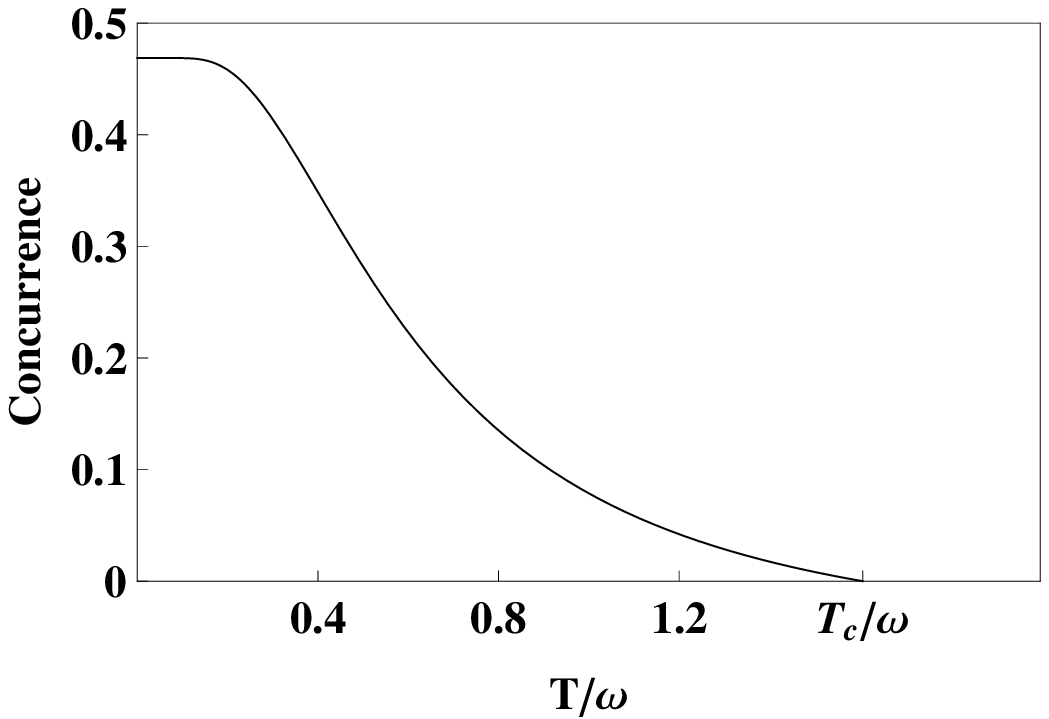}}\caption{$C_{\mr{as}}(\al)$
as a function of temperature for the  state (\ref{puresep}) with
$\al=1/4$.}
\end{figure}
\par
As the example of mixed separable initial state, consider the Gibbs state
(\ref{Gibbs}) at some temperature $T_{0}\neq T$. Notice that the fidelity $F$ of that state is given by
\begin{figure}[h]
\centering
{\includegraphics[height=54mm]{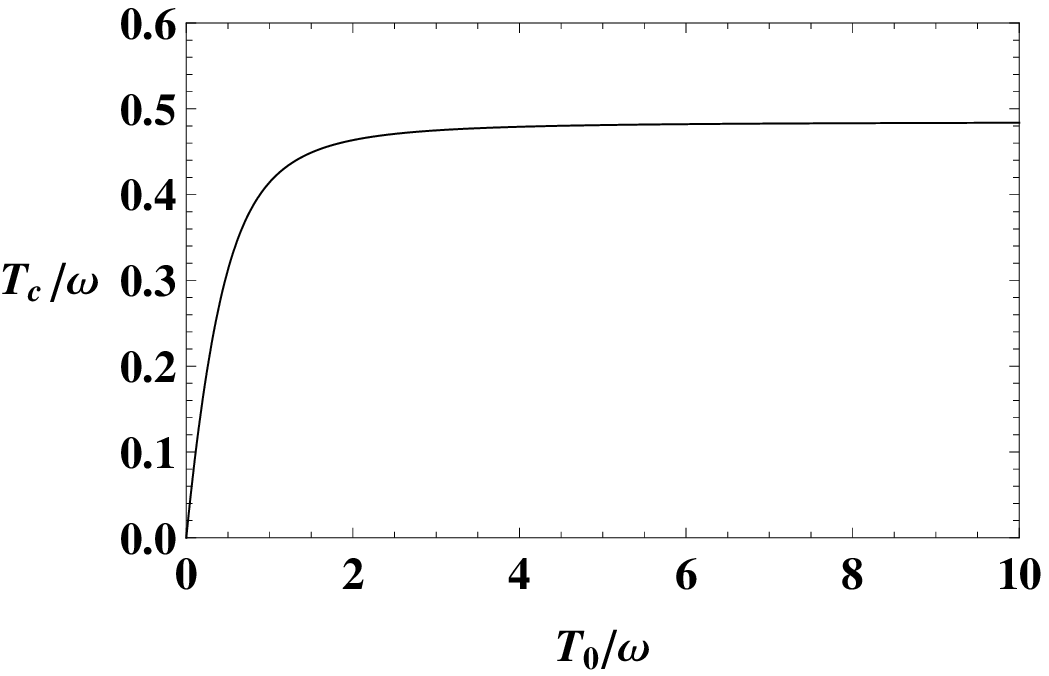}}\caption{Critical temperature versus $T_{0}$.}
\end{figure}
\begin{equation}
F=\frac{e^{-\be_{0}\om}}{(1+e^{-\be_{0}\om})^{2}},\quad \be_{0}=\frac{1}{T_{0}}\label{FGibbs}
\end{equation}
and the value of $F$ is always smaller then $1/2$. So there is the
critical temperature $T_{\mr{c}}$ of the reservoir for the creation
of asymptotic entanglement and $T_{\mr{c}}$ depends on $T_{0}$.
Explicit function describing $T_{\mr{c}}$ can be obtained by
combining the formulas (\ref{Tc}) and (\ref{FGibbs}). Since the
resulting function is complicated, we will not reproduce it here and
show only its plot (FIG. 3). Observe that the critical temperature
is always below the temperature $T_{0}$, thus in order to entangle
the atomic system prepared in the Gibbs state at temperature
$T_{0}$, the temperature of the photon reservoir have to be much
smaller then $T_{0}$. On the other hand, the amount of the created
entanglement can be computed from the formula (\ref{aconc}).
Similarly as in the case of pure initial states, asymptotic
concurrence is a decreasing function of temperature and vanishes at
$T_{\mr{c}}$ (FIG. 4).
\begin{figure}[h]
\centering
{\includegraphics[height=54mm]{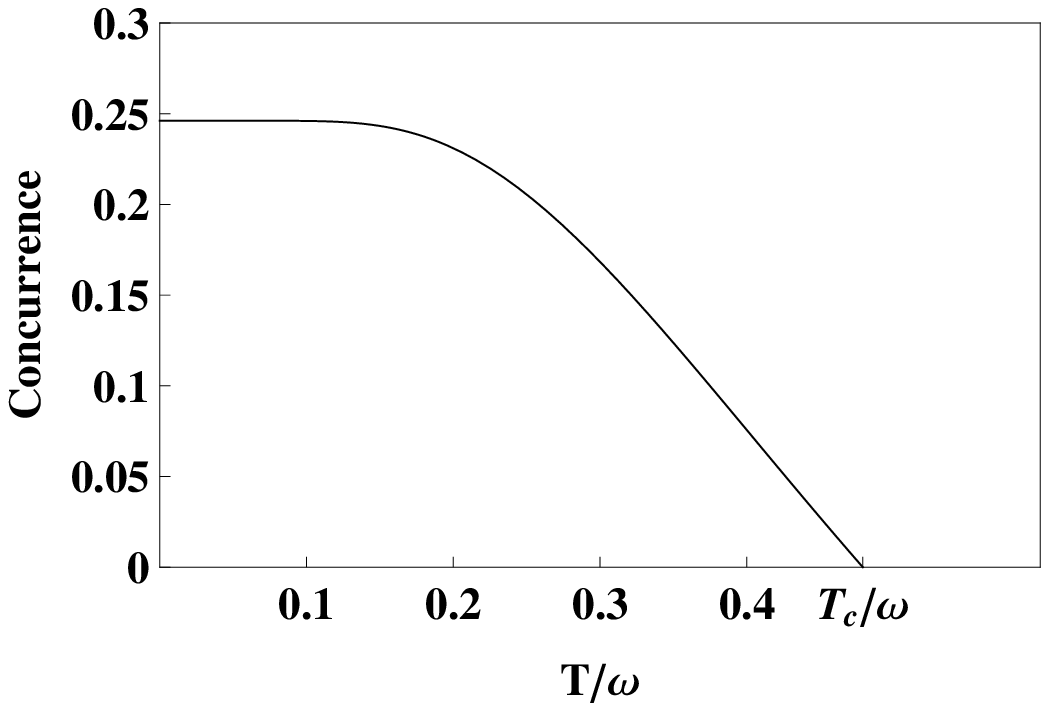}}\caption{$C_{\mr{as}}$ as a
function of temperature for the Gibbs state with $T_{0}/\om=4$.}
\end{figure}
\subsection{Entangled initial states}
The collective states $\ket{s}$ and $\ket{a}$ are specific instances
of states with maximal possible entanglement. In the case of two
qubits the set of maximally entangled states form a three-parameter
family of states. As was shown in \cite{BJO}, the corresponding
projectors can be parametrized as follows
$$
P_{a,\te_{1},\te_{2}}=
$$
\begin{equation}
\frac{1}{2}\,\begin{pmatrix}
a^{2}&ce^{-i\te_{1}}&ce^{-i\te_{2}}&-a^{2}e^{-i(\te_{1}+\te_{2})}\\[1mm]
ce^{i\te_{1}}&b^{2}&b^{2}e^{i(\te_{1}-\te_{2})}&-ce^{-i\te_{2}}\\[1mm]
ce^{i\te_{2}}&b^{2}e^{-i(\te_{1}-\te_{2})}&b^{2}&-ce^{-\te_{1}}\\[1mm]
-a^{2}e^{i(\te_{1}+\te_{2})}&-ce^{i\te_{2}}&-ce^{i\te_{1}}&a^{2}
\end{pmatrix}\label{maxent}
\end{equation}
where $b^{2}=1-a^{2},\; c=a\sqrt{1-a^{2}}$ and
$$
a\in [0,1],\quad \te_{1},\te_{2}\in [0,2\pi]
$$
All the states (\ref{maxent}) have concurrence equal to $1$, but
\begin{equation}
F=\frac{1}{2}\,(1-a^{2})\left(\,1-\cos(\te_{1}-\te_{2})\,\right)\label{Fmax}
\end{equation}
Notice that the fidelity $F$ can take all values from $0$ to $1$, depending on
parameters $a$ and $\te=\te_{1}-\te_{2}$.
In particular $F>1/2$ inside the set $\mathcal E$ on the $(a,\te)$ plane, given by
\begin{equation}
\mathcal E=\{\,0\leq a\leq\frac{1}{\sqrt{2}},\; \arccos \frac{a^{2}}{a^{2}-1}
\leq \te\leq 2\pi-\arccos\frac{a^{2}}{a^{2}-1}\,\}
\label{E}
\end{equation}
On the boundary of ${\mathcal E},\; F=1/2$ and outside this set,
$F<1/2$. So all initial states (\ref{maxent}) with
$(a,\te_{1}-\te_{2})\in {\mathcal E}$ remain entangled
asymptotically, for any finite temperature of the reservoir. The
asymptotic concurrence can be computed from the formula
\begin{equation}
C_{\mr{as}}(a,\te)=\frac{2(1-a^{2})\left(\sin^{2}\frac{\DS\te}{\DS
2}\cosh\bom-\cos\te\,\right)-(1+2a^{2})}{1+2\cosh\bom}
\label{maxconc}
\end{equation}
for $(a,\te)\in \mathcal E$. Notice that for all such initial states
the value of asymptotic concurrence is smaller then $1$, except
antisymmetric collective state $\ket{a}$ which is stable during the
evolution. As in the previous cases, the asymptotic concurrence is
maximal for the reservoir at zero temperature, with the value given
by
\begin{equation}
C_{\mr{max}}=(1-a^{2})\,\sin^{2}\frac{\DS \te}{\DS 2},
\end{equation}
but in the present case, it decreases to the nonzero minimal value
$C_{\mr{min}}$ when the temperature of the reservoir increases to
infinity, and
\begin{equation}
C_{\mr{min}}=-\,\left(a^{2}+(1-a^{2})\cos\te\,\right)
\end{equation}
\par
For the parameters $(a,\te)$ lying outside the set $\mathcal E$,
there is  the critical temperature $T_{\mr{c}}$ given by (\ref{Tc})
and (\ref{fTc}) for fidelity (\ref{Fmax}). For $T<T_{\mr{c}}$, the
interaction with reservoir preserves some initial entanglement, but
for $T\geq T_{\mr{c}}$, all maximally entangled states with such
parameters $a$ and $\te$ disentangle asymptotically. Some initial
states (\ref{maxent}) disentangle exactly to the equilibrium Gibbs
states at the specific temperature $\widetilde{T}$, depending on $a$
and $\te$. This interesting phenomenon  can happen when
\begin{equation}
(1-a^{2})\,\left(\,1-\cos\te\,\right)=\frac{e^{-\widetilde{\be}\om}}
{(1+e^{-\widetilde{\be}\om})^{2}},\quad
\widetilde{\be}=\frac{1}{\widetilde{T}}\label{FG}
\end{equation}
This equation can be satisfied for some finite temperature
$\widetilde{T}$ only if the parameters $(a,\te)$ lie outside the
curve
\begin{equation}
\te=\arccos\frac{2a^{2}-1}{2\,(a^{2}-1)},\quad a\in
[0,\sqrt{3}/2]\label{curve}
\end{equation}
and then
\begin{equation}
\widetilde{T}=\left[\,\ln\frac{a^{2}+(1-a^{2})\cos\te+
\sqrt{2a^{2}-1+2(1-a^{2})\cos\te}}{(1-a^{2})(1-\cos\te)}\,\right]^{-1}
\label{tildetemp}
\end{equation}
On the curve (\ref{curve}), this temperature is infinite.
\par
When the initial states are  non-maximally entangled, the
interaction with the photon reservoir can in some cases increase the
initial entanglement. To show that this is possible, consider the
class of states
\begin{equation}
\ro=\begin{pmatrix} 0&\hspace*{3mm}0&\hspace*{3mm}0&\hspace*{3mm}0\\[2mm]
0&\hspace*{3mm}x&-\frac{\DS z}{\DS 2}&\hspace*{3mm}0\\[2mm]
0&-\frac{\DS z}{\DS 2}&1-x&\hspace*{3mm}0\\[2mm]
0&\hspace*{3mm}0&\hspace*{3mm}0&\hspace*{3mm}0
\end{pmatrix}\label{inistates}
\end{equation}
where $x,z\in (0,1)$ are related by inequality
$$
\frac{z^{2}}{4}\leq x(1-x)
$$
Note that as the explicit examples of states (\ref{inistates}) we
can consider the pure states
\begin{equation}
\ket{\Psi}_{\eta}=\cos\eta \;\ket{0}_{A}\otimes\ket{1}_{B}+\sin\eta
\;\ket{1}_{A}\otimes\ket{0}_{B}\label{inipure}
\end{equation}
with $\eta\in (\pi/2,\pi)$. The states (\ref{inistates}) are
entangled with concurrence equal to $z$, and
$$
F=\frac{1+z}{2}>\frac{1}{2}
$$
\begin{figure}[t]
\centering
{\includegraphics[height=54mm]{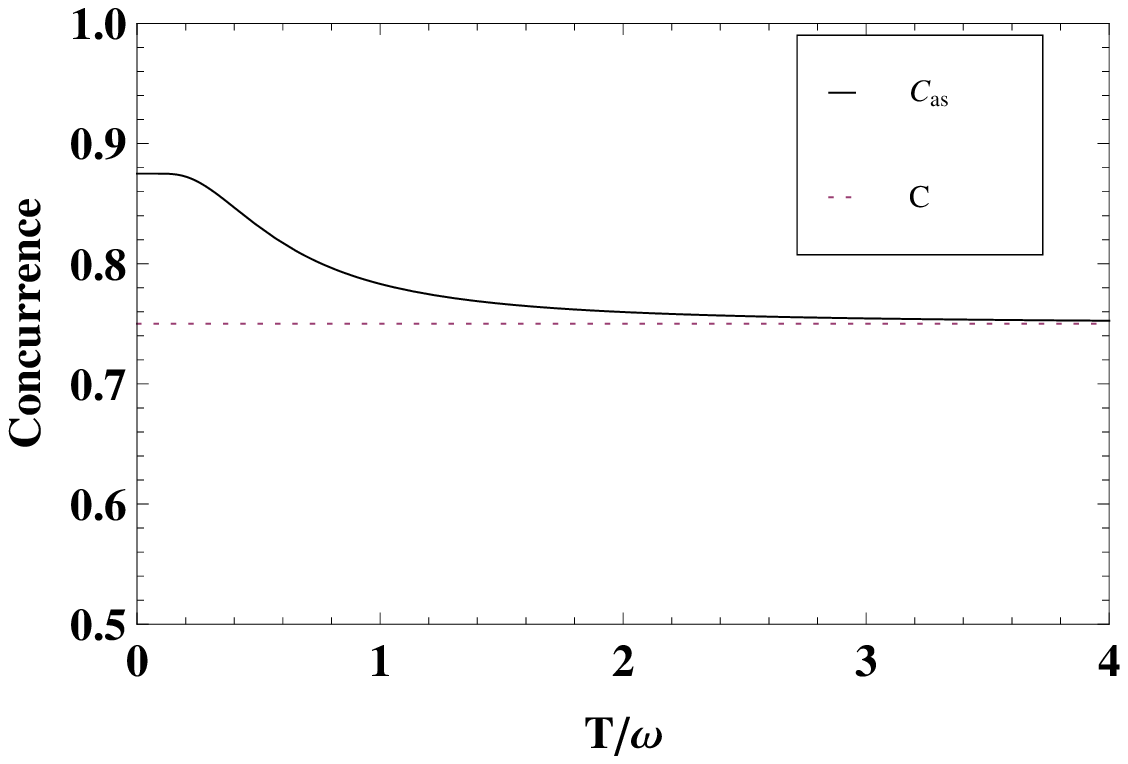}}\caption{Asymptotic entanglement as a function of temperature
for the initial state (\ref{inistates}). Here $C$ is the initial entanglement.}
\end{figure}
For every temperature of the reservoir the asymptotic concurrence is
non-zero and is given by
\begin{equation}
C_{\mr{as}}(z)=z+\frac{(1-z)(\cosh\bom-1)}{1+2\cosh\bom}\label{nonmaxconc}
\end{equation}
So it is greater then the initial concurrence  and the maximal
production of the additional entanglement happens at zero
temperature (FIG. 5). In that case
$$
C_{\mr{max}}(z)=z+\frac{1-z}{2}
$$
Observe also that when $T\to \infty$, $ C_{\mr{as}}(z)\to z $, so in
the reservoir at infinite temperature the initial entanglement is
exactly preserved \cite{JJ}.
\section{Conclusions}
We have investigated the dynamics of two-level atoms immersed in the
photon reservoir at finite temperature $T$. In the regime of strong
correlations between the atoms, there are nontrivial stationary
asymptotic states which are parametrized by the fidelity $F$ i.e.
the overlap of the initial state of the atoms with the singlet state
and the temperature $T$. For the values of $F$ above the threshold
fidelity, these states can be identified with thermal Werner states,
which are natural generalizations of standard Werner states.
Depending on $F$ and $T$, the asymptotic states can be separable or
entangled. Thus the dynamics describes the process of creation of
entanglement or the phenomenon of disentanglement. Concerning
generation of entanglement it is worth to stress that there exists a
critical temperature  above which the entanglement cannot be
created. Besides, even in the case of generation of entanglement,
the temperature diminishes its production. The maximal value of
entanglement is obtained for the case of zero temperature. On the
other hand, in the process of disentanglement some part of initial
entanglement can be preserved but we can also find such entangled
initial states that disentangle exactly to the Gibbs equilibrium
state.

\end{document}